\documentclass{article}
\usepackage{makeidx}
\usepackage{graphicx}
\usepackage{amsmath}

\begin{document}

\begin{flushright}
CGPG-00/7-2; WATPHYS-TH00/03
\end{flushright}



\vspace{1cm}

\begin{center}
{\Large \textbf{Conserved Quantities in Kerr-anti-de Sitter Spacetimes in
Various Dimensions}} 

\vspace{1cm}

Saurya Das {\footnote{%
EMail: das@gravity.phys.psu.edu} } \\[0pt]
Center for Gravitational Physics and Geometry, Department of Physics, \\[0pt]
The Pennsylvania State University, University Park, PA 16802-6300, USA\\[0pt]

\vspace{.5cm}

Robert B. Mann {\footnote{%
EMail: mann@avatar.uwaterloo.ca}} \\[0pt]
Department of Physics, University of Waterloo, \\[0pt]
Waterloo, Ontario N2L 3G1, CANADA
\end{center}

\vspace{2cm}

\begin{center}
ABSTRACT
\end{center}

We compute the conserved charges for Kerr anti-de Sitter spacetimes in
various dimensions using the conformal and the counterterm prescriptions. We
show that the conserved charge corresponding to the global timelike killing
vector computed by the two methods differ by a constant dependent on the
rotation parameter and cosmological constant in odd spacetime dimensions,
whereas the charge corresponding to the rotational killing vector is the
same in either approach. We comment on possible implications of our results
to the AdS/CFT correspondence.


\newpage

\section{Introduction}

Asymptotically anti-de Sitter (AAdS) spacetimes have attracted a great deal
of attention recently due to the conjectured AdS/CFT correspondence, which
relates supergravity/string theory in bulk AdS spacetimes to conformal field
theories on the boundary, the hope being that a full quantum theory of
gravity in AdS spacetimes can be replaced by a well understood
CFT/Yang-Mills theory, and observable quantities in the gravity theory can
be computed using the latter. A dictionary translating between different
quantities in the bulk gravity theory and its counterparts on the boundary
has emerged, including the partition functions and correlation functions of
both theories.

One of the fundamental set of quantities for any physical theory is the set
of conserved quantities associated with it. For theories of gravity on
asymptotically flat spacetimes, these are precisely the ADM conserved
quantities, which constitute the $d(d+1)/2$ conserved charges corresponding
to the Poincare generators in $d$-dimensions. However, the ADM formulae
break down for spacetimes that are AAdS (i.e. satisfying the Einstein
equations with a negative cosmological constant at timelike infinity),
implying that a new set of rules have to be laid down to construct conserved
quantities corresponding to the asymptotic $AdS_{d}$ group of isometries.
Efforts in this direction were made in \cite{by} \cite{bcm} . However, the
conserved charges constructed had supertranslation-like ambiguities (due to
the coordinate dependence of the formalism), or relied on using an auxiliary
spacetime, in which the boundary of the AAdS spacetime had to be embedded in
a reference spacetime. The latter procedure is neither unique, nor always
possible. In \cite{ashmag}, the Penrose conformal completion was used to
define conserved quantities for $d=4$, which removed the aforementioned
drawbacks. Its generalization to dimensions $d \geq 4$ was done in \cite
{ashdas}.

Independently, the AdS/CFT correspondence inspired an alternative approach
of constructing conserved quantities for AAdS spacetimes \cite{balakraus}.
This `counterterm' method proposes certain boundary terms (or counterterms)
which depend on the intrinsic geometry of the (timelike) boundary at large
spatial distances. These do not affect the bulk equations of motion and
eliminate the need to embed the given geometry in a reference spacetime.
However, this approach still involves taking the the problematic limit $%
r\rightarrow \infty $, where metric components necessarily diverge, unlike
the asymptotically flat spacetimes.

Although the above approaches can in principle be used to construct
conserved quantities for arbitrary AAdS spacetimes, in practice the
conformal formalism has only been used to compute the conserved quantity
associated with the global timelike killing vector field (KVF) for
Schwarzschild-AdS spacetimes for $d=4${\ and }$5$ (we will call this
quantity the mass, in analogy with the its counter part for asymptotically
flat geometries, although the group is no longer Poincare). On the other
hand, the counterterm method was used to calculate the conserved quantities
for Schwarzschild as well as Kerr-AdS \ (KAdS) geometries in four dimensions
\cite{rbkerr} \ as well as for $d\leq 7$ \cite{d=345}

In this paper we carry out a systematic analysis of KAdS black holes in
arbitrary dimensions $d\geq 4$ using the conformal as well as the
counterterm formalisms. We employ the algorithm of Kraus, Larsen and
Siebelink (KLS) \cite{kls} to construct the boundary counterterm
contributions up to $d=9$, and from this derive the action, mass and angular
momentum for $4\leq d\leq 9.$ We will show that for $d=$ odd, the mass of
the spacetime depends on which formalism we use and their difference is the
so called `Casimir energy', which is a function of $\Lambda $, the
cosmological constant and $a$, the angular momentum parameter. On the other
hand, the angular momentum of these solutions is independent of the method
of computation, and hence unambiguous. In either case, our results are
commensurate with the Gibbs-Duhem relation in semiclassical Euclidean
quantum gravity, which relates one-quarter of the area of the event horizon
to the minus the difference between the (Euclidean)\ action and the
Hamiltonian times the inverse temperature.

\section{The Conformal and Counterterm Methods}

Let us begin with a brief review of the methods under consideration. In the
conformal method, one begins with the following assumptions regarding the
physical $d$ dimensional AAdS manifold $\hat{M}$ with metric ${\hat{g}}_{ab}$
and the conformally mapped manifold $M$ with metric $g_{ab}$:

\noindent (i) $g_{ab}=\Omega ^{2}~{\hat{g}}_{ab}$, where $\Omega $ is a
non-negative conformal factor. \newline
(ii) the boundary $\mathcal{I}$ of $M$ is topologically $S^{(d-2)}\times R$,
$\Omega $ vanishes on $\mathcal{I}$, but $\nabla _{a}\Omega \neq 0$ on $%
\mathcal{I}$. \newline
(iii) The Einstein equations with $\Lambda <0$ are satisfied on $\mathcal{I}
$. \newline
(iv) The fall-off behaviors of the matter fields and the Weyl tensor are
such that $\Omega ^{2-d}T_{ab}$ and $\Omega ^{4-d}C_{abcd}$ are smooth on $%
\mathcal{I}$.

It can be easily verified that the $d$-dimensional KAdS spacetime satisfies
the above conditions. Then transforming all tensors to the conformally
mapped spacetime, and after a series of straightforward analyses, it can be
shown that the following equation is valid on $\mathcal{I}$:
\begin{equation*}
D^{p}\mathcal{E}_{mp}=\,-{8\pi}\,\,\frac{(d-3)}{\ell }\,\,\mathbf{T}%
_{ab}n^{a}h^{b}{}_{m}\,.
\end{equation*}
where $D^{d}$ is the intrinsic covariant derivative on $\mathcal{I}$,
compatible with the induced metric $h_{ab}:=g_{ab}-{\ell}%
^{2}n_{a}n_{b}~~(n_{a}:=\nabla _{a}\Omega )$ and $\mathcal{E}_{ab}$ is the
electric part of the Weyl tensor at $\mathcal{I}$ defined as: $\mathcal{E}%
_{ab}:={\ell}^{2}~\Omega ^{3-d}C_{ambn}n^{m}n^{n}$ (we have set the Newton's
constant to be unity here and in the following analyses). We have
parametrized the cosmological constant as $\Lambda =-(d-1)(d-2)/2{\ell}^{2}$
and $\mathbf{T}_{ab}:=\Omega ^{2-d}{\hat{T}}_{ab}$ on $\mathcal{I}$. {}From
the above equation, the conserved charge associated with the conformal KVF $%
\xi $ is defined as follows (note that an ordinary KVF on ${\hat M}$ becomes
the conformal KVF on $M$) :
\begin{equation}
Q_{\xi }[C]:=-\frac{1}{8\pi}~\frac{\ell }{d-3}~~\oint_{C}\mathcal{E}_{ab}\xi
^{a}dS^{b}  \label{conserve}
\end{equation}
which satisfies the balance equation
\begin{equation}
Q_{\xi }[C_{2}]-Q_{\xi }[C_{1}]=\int_{\Delta \mathcal{I}}\mathbf{T}_{ab}\xi
^{a}dS^{b}  \label{balance}
\end{equation}
in presence of matter fields ($C_{1}$ and $C_{2}$ are two cross sections on $%
\mathcal{I}$ bounding the region $\Delta \mathcal{I}$). Equations (\ref
{conserve}) and (\ref{balance}) are the fundamental relations which we will
use to define conserve quantities.

Now we write down the explicit KAdS family of solutions. For simplicity we
will assume only one rotation parameter \cite{hawk}:
\begin{eqnarray}
ds^{2} &=&-\frac{\Delta _{r}}{\rho ^{2}}(dt-\frac{a}{\Xi }\sin ^{2}\theta
d\phi )^{2}+\frac{\rho ^{2}}{\Delta _{r}}dr^{2}+\frac{\rho ^{2}}{\Delta
_{\theta }}d\theta ^{2}  \label{kadsmet} \\
&&+\frac{\Delta _{\theta }\sin ^{2}\theta }{\rho ^{2}}[adt-\frac{%
(r^{2}+a^{2})}{\Xi }d\phi ]^{2}+r^{2}\cos ^{2}\theta d\Omega _{d-4}^{2}
\notag
\end{eqnarray}
where
\begin{eqnarray}
\Delta _{r} &=&(r^{2}+a^{2})(1+r^{2}/l^{2})-2mr^{5-d};  \notag \\
\Delta _{\theta } &=&1-a^{2}\cos ^{2}\theta /l^{2};  \label{kadsnotn} \\
\Xi &=&1-a^{2}/l^{2};  \notag \\
\rho ^{2} &=&r^{2}+a^{2}\cos ^{2}\theta .  \notag
\end{eqnarray}
and $d\Omega _{d-4}^{2}$ is the line element on unit $S^{d-4}$. The
parameters $m$ and $a$ are related to the mass and angular momentum of the
black hole, as we shall see.

Evaluation of the electric part of the Weyl tensor (at $\mathcal{I}$) yields
to leading order in $r$:

\begin{center}
\begin{tabular}{ccc}
\multicolumn{3}{c}{Table I$~~~~~~~~~~~$} \\
$\text{{Dim}}$ & $\mathcal{\hat{E}}_{tt}$ & $~~\mathcal{\hat{E}}_{t\phi }$
\\
\multicolumn{1}{l}{$~~4$} & \multicolumn{1}{l}{$-2m/l^{2}r$} &
\multicolumn{1}{l}{$2\sin ^{2}\theta ~m~a/\Xi ~l^{2}r$} \\
\multicolumn{1}{l}{$~~5$} & \multicolumn{1}{l}{$-6m/l^{2}r^{2}$} &
\multicolumn{1}{l}{$6\sin ^{2}\theta ~m~a/\Xi ~l^{2}r^{2}$} \\
\multicolumn{1}{l}{$~~6$} & \multicolumn{1}{l}{$-12m/l^{2}r^{3}$} &
\multicolumn{1}{l}{$12\sin ^{2}\theta ~m~a/\Xi ~l^{2}r^{3}$} \\
\multicolumn{1}{l}{$~~7$} & \multicolumn{1}{l}{$-20m/l^{2}r^{4}$} &
\multicolumn{1}{l}{$20\sin ^{2}\theta ~m~a/\Xi ~l^{2}r^{4}$} \\
\multicolumn{1}{l}{$~~8$} & \multicolumn{1}{l}{$-30m/l^{2}r^{5}$} &
\multicolumn{1}{l}{$30\sin ^{2}\theta ~m~a/\Xi ~l^{2}r^{5}$} \\
\multicolumn{1}{l}{$~~9$} & \multicolumn{1}{l}{$-42m/l^{2}r^{6}$} &
\multicolumn{1}{l}{$42\sin ^{2}\theta ~m~a/\Xi ~l^{2}r^{6}$}
\end{tabular}
\index{table1}
\end{center}

\bigskip

To evaluate the surface element $dS^{b}$, we use the determinant of the
(hypothetical) induced metric on $C$:
\begin{equation}
\sigma _{ab}:=g_{ab}-~l^{2}n_{a}n_{b}+u_{a}u_{b}
\end{equation}
where $u^{a}$ is the timelike unit normal at $C$.

Finally, using the timelike KVF $\partial /\partial t$ and the rotational
KVF $\partial /\partial \phi $, and transforming back the integrals at $%
\mathcal{I}$ to the physical space time ${%
\hat{M}}$, the mass and angular momenta of the KAdS spacetime in various
dimensions can be calculated. The results are presented in table II.

\bigskip

Now we move on to the counterterm action and the conserved charges obtained
thereof. As is well known, the Einstein action with a negative cosmological
constant along with the Gibbons-Hawking boundary term (in the remaining
analysis we will always work in the physical spacetime, and omit the hats
over the geometrical quantities, for brevity)
\begin{equation}
I=\frac{1}{8\pi G_{(d)}}\left( \int_{M}d^{d}x\sqrt{-g}\left[ R-2\Lambda %
\right] -\int_{\partial M}~d^{d-1}x\sqrt{-\gamma }K\left( \gamma \right)
\right)  \label{act}
\end{equation}
(where $K$ is the trace of the extrinsic curvature of the timelike boundary
and $\gamma_{ab}$ is the induced metric on this boundary) is divergent for
the spacetimes under consideration, with fall off conditions natural to this
setting. To eliminate this divergence, the KLS counterterm proposal
prescribes adding terms to the action which are \textit{intrinsic} to the
boundary as $r\rightarrow \infty $. These take the form \cite
{balakraus,rbkerr,d=345,kls}
\begin{equation}
I_{ct}=\frac{1}{8\pi G_{(d)}}\int_{\partial M}\mathcal{L}_{\text{ct}}
\label{actct}
\end{equation}
where the quantity
\begin{equation}
\tilde{\Pi}_{ab}\equiv \frac{2}{\sqrt{-\gamma }}\frac{\delta S_{ct}}{\delta
\gamma ^{ab}}  \label{Pict}
\end{equation}
is a solution to the Gauss-Coddacci relations
\begin{equation}
\frac{1}{d-2}\tilde{\Pi}^{2}-\tilde{\Pi}_{ab}\tilde{\Pi}^{ab}=\frac{\left(
d-1\right) \left( d-2\right) }{\ell ^{2}}+R  \label{Gc}
\end{equation}
as a power series in $1/\ell $, i.e. $\tilde{\Pi}_{ab}=\frac{1}{\ell }%
\sum_{n=0}$\ $\ell ^{2n}$ $\tilde{\Pi}_{ab}^{(n)}$. To order $n$, the
relation (\ref{Gc}) is linear in the trace $\tilde{\Pi}^{(n)}$, and so this
quantity may be determined in terms of the lower-order terms. Under local
Weyl rescalings, the relation (\ref{Pict}) ensures that
\begin{equation}
\mathcal{L}_{\text{ct}}^{(n)}=\frac{\tilde{\Pi}^{(n)}}{d-1-2n}
\label{Lct-trace}
\end{equation}
and by then by using (\ref{Pict}) again, the full expression for $\tilde{\Pi}%
_{ab}$ can be obtained to the desired order. \

This procedure yields
\begin{eqnarray}
&{}& \mathcal{L}_{\text{ct}} = -\frac{d-2}{\ell }\sqrt{-\gamma }-\frac{\ell
\sqrt{-\gamma }}{2(d-3)}R-\frac{\ell ^{3}\sqrt{-\gamma }}{2(d-3)^{2}(d-5)}%
\left( R^{ab}R_{ab}-\frac{d-1}{4(d-2)}R^{2}\right)  \notag \\
&+&\frac{\ell ^{5}\sqrt{-\gamma }}{(d-3)^{3}(d-5)(d-7)}\left( \frac{3d-1}{%
4(d-2)}RR^{ab}R_{ab}-\frac{\left( d-1\right) (d+1)}{16(d-2)^{2}}R^{3} \right.
\notag \\
&-& 2R^{ab}R^{cd}R_{acbd} + \left. \frac{d-3}{2(d-2)}R^{ab}\nabla _{a}\nabla
_{b}R-R^{ab}\nabla ^{2}R_{ab}+\frac{1}{2(d-2)}R\nabla ^{2}R\right)
\label{Lct}
\end{eqnarray}
to order $\ell ^{5}$. All the geometrical quantities above are intrinsic to
the timelike boundary at $r \rightarrow \infty$). An integration by parts
renders this expression in the more convenient form
\begin{eqnarray}
\mathcal{L}_{\text{ct}} &=&-\frac{d-2}{\ell }\sqrt{-\gamma }-\frac{\ell
\sqrt{-\gamma }}{2(d-3)}R-\frac{\ell ^{3}\sqrt{-\gamma }}{2(d-3)^{2}(d-5)}%
\left( R^{ab}R_{ab}-\frac{d-1}{4(d-2)}R^{2}\right)  \notag \\
&&+\frac{\ell ^{5}\sqrt{-\gamma }}{(d-3)^{3}(d-5)(d-7)}\left( \frac{3d-1}{%
4(d-2)}RR^{ab}R_{ab}-\frac{\left( d-1\right) (d+1)}{16(d-2)^{2}}R^{3} \right.
\notag \\
&&\text{ \ \ \ \ \ \ \ }\left. -2R^{ab}R^{cd}R_{acbd} -\frac{d-1}{4(d-2)}%
\nabla _{a}R\nabla ^{a}R+\nabla ^{c}R^{ab}\nabla _{c}R_{ab}\right)
\label{Lct2}
\end{eqnarray}
Varying the action with respect to the boundary metric $\gamma _{ab}$, the
full stress-energy tensor for gravity is defined as:
\begin{equation}
T_{ab}:=\frac{2}{\sqrt{-\gamma }}\frac{\delta }{\delta \gamma ^{ab}}\left(
S+S_{\text{ct}}\right)  \label{Tab}
\end{equation}
which results in the boundary stress-energy:

\begin{eqnarray}
&{}&T_{ab}=K_{ab}-\gamma _{ab}K-\frac{d-2}{l}\gamma _{ab}+\frac{l}{d-3}%
\left( R_{ab}-\frac{1}{2}\gamma _{ab}R\right)  \notag \\
&&+\frac{l^{3}}{(d-3)^{2}(d-5)}\left\{ -\frac{1}{2}\gamma _{ab}\left(
R_{cd}R^{cd}-\frac{(d-1)}{4(d-2)}R^{2}\right) -\frac{(d-1)}{2(d-2)}%
RR_{ab}\right.  \notag \\
&&\left. +2R^{cd}R_{cadb}-\frac{d-3}{2(d-2)}\nabla _{a}\nabla _{b}R+\nabla
^{2}R_{ab}-\frac{1}{2(d-2)}\gamma _{ab}\nabla ^{2}R\right\}  \notag \\
&& - \frac{2\ell ^{5}\sqrt{-\gamma }}{(d-3)^{3}(d-5)(d-7)}\left\{ \frac{3d-1}{%
4(d-2)}\left[ \left( G_{ab}R^{cd}R_{cd} \right) \right. \right.  \notag \\
&& \left. \left. -\nabla _{a}\nabla _{b}\left( R^{ef}R_{ef}\right) +\gamma
_{ab}\nabla ^{2}\left( R^{ef}R_{ef}\right) \right. \right.  \notag \\
&&\left. +2RR_{a}^{\;c}R_{bc}+\gamma _{ab}\nabla _{c}\nabla _{d}\left(
RR^{cd}\right) +\nabla ^{2}\left( RR_{ab}\right) -\nabla ^{c}\nabla
_{b}\left( RR_{ac}\right) -\nabla ^{c}\nabla _{a}\left( RR_{bc}\right)
\right]  \notag \\
&&-\frac{\left( d-1\right) (d+1)}{16(d-2)^{2}}\left[ -\frac{1}{2}\gamma
_{ab}R^{3}+3R^{2}R_{ab}-3\nabla _{a}\nabla _{b}R^{2}+3\gamma _{ab}\nabla
^{2}R^{2}\right]  \notag \\
&&-2\left[ -\frac{1}{2}\gamma _{ab}R^{ef}R^{cd}R_{ecfd}+\frac{3}{2}\left(
R_{a}^{\;e}R^{cd}R_{ecbd}+R_{b}^{\;e}R^{cd}R_{ecad}\right) \right.  \notag \\
&&\left. -\nabla_{c}\nabla _{d}\left( R_{ab}R^{cd}\right) +\nabla _{c}\nabla
_{d}\left( R_{a}^{\;c}R_{b}^{\;d}\right) \right.  \notag \\
&&\left. +\gamma _{ab}\nabla _{e}\nabla ^{f}\left(
R^{cd}R_{\;cfd}^{e}\right) +\nabla ^{2}\left( R^{cd}R_{acbd}\right) -\nabla
_{e}\nabla _{a}\left( R^{cd}R_{\;cbd}^{e}\right) -\nabla _{e}\nabla
_{b}\left( R^{cd}R_{\;cad}^{e}\right) \right]  \notag \\
&&-\frac{d-1}{4(d-2)}\left[ \nabla _{a}R\nabla _{b}R-\frac{1}{2}\gamma
_{ab}\left( \nabla _{c}R\nabla ^{c}R\right) -2R_{ab}\nabla ^{2}R-2\gamma
_{ab}\nabla ^{4}R+2\nabla _{a}\nabla _{b}\nabla ^{2}R\right]  \notag \\
&&+\left[ 2\nabla _{c}R_{ad}\nabla ^{c}R_{b}^{\;d}+\nabla _{a}R^{cd}\nabla
_{b}R_{cd}-\frac{1}{2}\gamma _{ab}\nabla ^{e}R^{cd}\nabla _{e}R_{cd}\right.
\notag \\
&&-\gamma _{ab}\nabla _{c}\nabla _{d}\nabla ^{2}R^{cd}-\nabla
^{4}R_{ab}+\nabla _{c}\nabla _{a}\nabla ^{2}R_{b}^{\;c}+\nabla _{c}\nabla
_{b}\nabla ^{2}R_{a}^{\;c}-\nabla _{c}\left( R_{bd}\nabla _{a}R^{cd}\right)
\notag \\
&&\left. -\nabla _{c}\left( R_{ad}\nabla _{b}R^{cd}\right) -\nabla
_{c}\left( R_{ad}\nabla ^{c}R_{b}^{\;d}+R_{bd}\nabla ^{c}R_{a}^{\;d}\right)
\right.  \notag \\
&& \left. +\nabla ^{c}\left( R_{cd}\nabla _{a}R_{b}^{\;d}\right) +\nabla
^{c}\left( R_{cd}\nabla _{b}R_{a}^{\;d}\right) \right]  \label{TabCFT}
\end{eqnarray}
which is valid for all $4\leq d\leq 9$. \ For a given dimension $d$, all
terms of order greater than $\ell ^{\left[ \frac{d}{2}+1\right] }$ give
vanishing contributions to the conserved charges at infinity, where $\left[
\frac{n}{2}\right] $ is the largest integer less than $\frac{n}{2}$. We do a
case by case analysis to determine the relevant terms in any specific
spacetime dimension.

Here, apart from the intrinsic curvature terms, one evaluates the extrinsic
curvature $K_{ab}$ of the boundary. The final expression for the conserved
charge associated with an (ordinary) KVF $\xi $ is, in this case:
\begin{equation}
Q_{\xi }:=\int_{\sigma }d^{d-2}x\sqrt{\sigma }T_{ab}u^{a}\xi ^{b}
\label{QCFT}
\end{equation}
where $u^{a}$ is the timelike unit normal at the boundary cylinder at $%
r\rightarrow \infty $ and $\sigma _{ab}:=\gamma _{ab}-u_{a}u_{b}$.

We have computed from (\ref{TabCFT}) the components of $T_{ab}$ for $%
d=4,5,6,7,8,9$. Our results for $d\leq 7$ are in agreement with previous
results \cite{rbkerr} \cite{d=345} \cite{d=9}; since the expression are
somewhat long and cumbersome (especially for the last two cases) we shall
not reproduce them here. \ Inserting these expressions into (\ref{QCFT}) for
the curvature tensors and the extrinsic curvatures, we finally obtain the
values of the timelike and rotational conserved charges for KAdS spacetimes
in various dimensions. The results also appear in table II, for easy
comparison with the conformal case. The masses computed from the conformal
method and that from the CFT method are given in units of $\pi /\Xi $ and
the angular momentum in units of $am\pi /\Xi ^{2}$. \ The angular momenta as
calculated with either method yield the same values, and so we have provided
only one column for this case. \ The masses differ for $d=$odd because of
the presence of the Casimir energy induced by the extra boundary terms.

\begin{center}
$\overset{}{
\begin{tabular}{llcl}
\multicolumn{4}{c}{Table\ II} \\
$\text{{Dim}}$ & $\text{~~Mass }$ & $\text{Mass}$ & $\text{Angular}$ \\
${}$ & $\text{(conformal)}$ & $\text{(counterterm)}$ & $\text{Momentum}$ \\
$~4$ & $~~\frac{m}{\pi }$ & $\frac{m}{\pi }$ & $~~~~1/\pi $ \\
$~5$ & $~~\frac{3m}{4}$ & $\frac{3}{4}m+\frac{9\ell ^{4}-9a^{2}\ell
^{2}+a^{4}}{96\ell ^{2}}$ & $~~~~1/2$ \\
$~6$ & $~~\frac{4m}{3}$ & $\frac{4m}{3}$ & $~~~~2/3$ \\
$~7$ & $~~\frac{5m\pi }{8}$ & $\frac{5\pi m}{8}-\frac{\pi (a^{6}+5a^{4}\ell
^{2}-50a^{2}\ell ^{4}+50\ell ^{6})}{1280\ell ^{2}}$ & $~~~~\pi /4$ \\
$~8$ & $~~\frac{4\pi m}{5}$ & $\frac{4\pi m}{5}$ & $~~~~4\pi /15$ \\
$~9$ & $~~\frac{7m\pi ^{2}}{24}$ & $\frac{7m\pi ^{2}}{24}-\frac{\pi
^{2}(3a^{8}-41a^{6}\ell ^{2}-87a^{4}\ell ^{4}+1225a^{2}\ell ^{6}-1225\ell
^{8})}{107520\ell ^{2}}$ & $~~~~\pi ^{2}/12$%
\end{tabular}
}$
\end{center}

We have also cross-checked these results with the Gibbs-Duhem relation
\cite{GDrel}, which states that
\begin{equation}
S=\beta H_{\infty }-I  \label{gd}
\end{equation}
where $I$ is the Euclidean space action and $H_{\infty }=M-\Omega J$, where $%
M$ and $J$\ are respectively the mass and angular momentum as computed by
either method and $\Omega =\frac{a\Xi }{r_{+}^{2}+a^{2}}$ is the angular
velocity at the horizon for all dimensionalities. \ The entropy $S$ is given
by one-quarter of the horizon area for the KAdS spacetimes. \ In table III
we list the results of a computation of the action using equations (\ref
{actct},\ref{Lct}), along with the values for the inverse temperature and
entropy. \ The boxed terms for each of the odd-dimensional cases are the
extra terms in the action induced by the counterterm contributions; these
terms would are absent if we compute the action using the methods of refs.
\cite{by},\cite{bcm}.

\begin{center}
\begin{tabular}{lccc}
\multicolumn{4}{c}{Table III} \\
$d$ & $
\begin{array}{l}
\text{Inverse} \\
\text{Temperature}
\end{array}
$ & Entropy & Action \\
$4$ & \multicolumn{1}{l}{$\frac{4\pi \ell ^{2}r_{+}\left(
r_{+}^{2}+a^{2}\right) }{\left( r_{+}^{2}\ell
^{2}+3r_{+}^{4}+a^{2}r_{+}^{2}-a^{2}\ell ^{2}\right) }$} &
\multicolumn{1}{l}{$\frac{\pi \left( r_{+}^{2}+a^{2}\right) }{\Xi }$} &
\multicolumn{1}{l}{$-\frac{\pi \left( r_{+}^{2}+a^{2}\right) ^{2}\left(
r_{+}^{2}-\ell ^{2}\right) }{\Xi \left( r_{+}^{2}\ell
^{2}+3r_{+}^{4}+a^{2}r_{+}^{2}-a^{2}\ell ^{2}\right) }$} \\
$5$ & \multicolumn{1}{l}{$\frac{2\pi \ell ^{2}\left( r_{+}^{2}+a^{2}\right)
}{r_{+}\left( \ell ^{2}+2r_{+}^{2}+a^{2}\right) }$} & \multicolumn{1}{l}{$%
\frac{\pi ^{2}r_{+}\left( r_{+}^{2}+a^{2}\right) }{2\Xi }$} &
\multicolumn{1}{l}{$-\frac{\pi ^{2}\left( r_{+}^{2}+a^{2}\right) }{48\Xi
r_{+}\left( \ell ^{2}+2r_{+}^{2}+a^{2}\right) } $} \\
& \multicolumn{1}{l}{} & \multicolumn{1}{l}{} & \multicolumn{1}{l}{$\times
\left(12\left(r_{+}^{2}+a^{2}\right) \left( r_{+}^{2}-\ell ^{2}\right)
\right. $} \\
& \multicolumn{1}{l}{} & \multicolumn{1}{l}{} & \multicolumn{1}{l}{$\left. +%
\left[ 9\ell ^{2}\left(a^{2}-\ell ^{2}\right) -a^{4}\right] \right) $} \\
$6$ & \multicolumn{1}{l}{$\frac{4\pi \ell ^{2}r_{+}\left(
r_{+}^{2}+a^{2}\right) }{\left( 3r_{+}^{2}\ell
^{2}+5r_{+}^{4}+3a^{2}r_{+}^{2}+a^{2}\ell ^{2}\right) }$} &
\multicolumn{1}{l}{$\frac{2\pi ^{2}r_{+}^{2}\left( r_{+}^{2}+a^{2}\right) }{%
3\Xi }$} & \multicolumn{1}{l}{$-\frac{2\pi ^{2}r_{+}^{2}\left(
r_{+}^{2}+a^{2}\right) ^{2}\left( r_{+}^{2}-\ell ^{2}\right) }{3\Xi \left(
3r_{+}^{2}\ell ^{2}+5r_{+}^{4}+3r_{+}^{2}a^{2}+a^{2}\ell ^{2}\right) }$} \\
$7$ & \multicolumn{1}{l}{$\frac{2\pi \ell ^{2}r_{+}\left(
r_{+}^{2}+a^{2}\right) }{\left( 2r_{+}^{2}\ell
^{2}+3r_{+}^{4}+2a^{2}r_{+}^{2}+a^{2}\ell ^{2}\right) }$} &
\multicolumn{1}{l}{$\frac{\pi ^{3}r_{+}^{3}\left( r_{+}^{2}+a^{2}\right) }{%
4\Xi }$} & \multicolumn{1}{l}{$-\frac{\pi ^{3}r_{+}\left(
r_{+}^{2}+a^{2}\right) }{640\Xi \left( 2r_{+}^{2}\ell
^{2}+3r_{+}^{4}+2r_{+}^{2}a^{2}+a^{2}\ell ^{2}\right) } $} \\
& \multicolumn{1}{l}{} & \multicolumn{1}{l}{} & \multicolumn{1}{l}{$\left.
\times \left(80r_{+}^{2}\left( r_{+}^{2}+a^{2}\right) \left( r_{+}^{2}-\ell
^{2}\right) \right. \right. $} \\
& \multicolumn{1}{l}{} & \multicolumn{1}{l}{} & \multicolumn{1}{l}{$\ \left.
-\left[ 50\ell ^{4}\left( a^{2}-\ell ^{2}\right) -a^{4}\left( a^{2}+5\ell
^{2}\right) \right] \right) $} \\
$8$ & \multicolumn{1}{l}{$\frac{4\pi \ell ^{2}r_{+}\left(
r_{+}^{2}+a^{2}\right) }{\left( 5r_{+}^{2}\ell
^{2}+7r_{+}^{4}+5a^{2}r_{+}^{2}+3a^{2}\ell ^{2}\right) }$} &
\multicolumn{1}{l}{$\frac{4\pi ^{3}r_{+}^{4}\left( r_{+}^{2}+a^{2}\right) }{%
15\Xi }$} & \multicolumn{1}{l}{$-\frac{4\pi ^{3}r_{+}^{4}\left(
r_{+}^{2}+a^{2}\right) ^{2}\left( r_{+}^{2}-\ell ^{2}\right) }{15\Xi \left(
5r_{+}^{2}\ell ^{2}+7r_{+}^{4}+5a^{2}r_{+}^{2}+3a^{2}\ell ^{2}\right) }$} \\
$9$ & \multicolumn{1}{l}{$\frac{2\pi \ell ^{2}r_{+}\left(
r_{+}^{2}+a^{2}\right) }{\left( 3r_{+}^{2}\ell
^{2}+4r_{+}^{4}+3a^{2}r_{+}^{2}+2a^{2}\ell ^{2}\right) }$} &
\multicolumn{1}{l}{$\frac{\pi ^{4}r_{+}^{5}\left( r_{+}^{2}+a^{2}\right) }{%
12\Xi }$} & \multicolumn{1}{l}{$-\frac{4\pi ^{4}\left(
r_{+}^{2}+a^{2}\right) r_{+}}{53760\Xi \left( 3r_{+}^{2}\ell
^{2}+4r_{+}^{4}+3r_{+}^{2}a^{2}+2a^{2}\ell ^{2}\right) } $} \\
&  &  & \multicolumn{1}{l}{$\times \left(2240r_{+}^{4}\left(
r_{+}^{2}+a^{2}\right) \left( r_{+}^{2}-\ell ^{2}\right) \right. $} \\
&  &  & \multicolumn{1}{l}{$\left. -\left[ 1225\ell ^{6}\left( a^{2}-\ell
^{2}\right)
\right. \right.
$} \\
&  &  & \multicolumn{1}{l}{$ \left. \left.
-a^{4}\left( 3a^{4}-41a^{6}\ell ^{2}-87\ell ^{4}\right)
\right] \right) $}
\end{tabular}
\end{center}

\section{Discussion}

As can be observed from table III, the expressions for both the mass and the
action obtained from the conformal and the counterterm prescriptions do not
agree in odd dimensionalities, whereas the angular momenta from the two
approaches agree for all dimensions. However, the disagreements are
precisely such that the extra (or Casimir) energies exactly balance the
additional action contributions and the Gibbs-Duhem relation is still
satisfied.

Several comments are in order here. The difference in mass for for $d=5$
calculated from the two different methods was interpreted as the Casimir
energy of $\mathcal{N}=4~~SU(N)$ Yang-Mills theory on the conformal boundary
$S^{3}\times R$ of $AdS_{5}$. It is not evident that the differences for $%
d=7~${and}$~9$ lend themselves to an analogous interpretation, because in
these cases, the boundary CFTs are not well understood. The additional
contribution to the mass is a decreasing positive function of $a/\ell $ over
the allowed range $\left| a\right| \leq \ell $ for $d=5,9$ whereas it is an
increasing negative function of $a/\ell $\ over this range for $d=7$, so the
CFT in the latter case must have a negative Casimir energy. \ Both methods
are consistent with the Gibbs-Duhem relation (\ref{gd}); in the counterterm
method, the additional contributions to the action from the counterterms are
exactly canceled by the Casimir contributions to the mass. \ None of the
additional contributions from the counterterms have a well-defined flat
space limit; both the mass and the action diverge as $\ell \rightarrow
\infty $. \

It is interesting to note that there is no corresponding `Casimir'
contribution for the rotational KVF. Perhaps a clearer understanding of this
is required in the light of the AdS/CFT correspondence. An interesting check
would be to compute the other conserved charges for multiple rotational
parameters for $d\geq 5$ and see whether there are Casimir like terms for
these charges. Note that from the purely general relativistic point of view,
any such terms can be ruled out by simple yet robust covariance arguments
\cite{ashdas}. It may be noted that the counterterms become more and more
complicated as the dimensionality of the spacetime increases, although they
can be uniquely fixed by requiring the elimination of divergences. On the
other hand, the conformal method fixes the expressions for the conserved
charges once and for all for all $d$ and their explicit computations boil
down to the computation of the Weyl curvature and the surface element on $%
\mathcal{I}$. An interesting exercise would be to calculate the conserved
charges using both the methods for more complicated AAdS spacetimes like
Taub-NUT-AdS and Taub-Bolt-AdS metrics. We hope to report on it elsewhere.

\bigskip

\begin{center}
\textbf{ACKNOWLEDGEMENTS}
\end{center}

We would like to thank A. Ashtekar for discussions. The work of S.D. was
supported by NSF grant NSF-PHY-9514240 and the Eberly research funds of Penn
State. R.B. Mann was supported by the Natural Sciences and Engineering
Research council of Canada.

\end{document}